\documentclass[twocolumn,prd,nofootinbib,aps,prl,floats,floatfix,amsmath,amssymb,longbibliography,secnumarabic]{revtex4-1} %

\usepackage{hyperref}
\usepackage{amsmath}
\usepackage{bbm}
\usepackage{amsfonts}
\usepackage{amssymb}
\usepackage{latexsym}
\usepackage{graphicx}
\usepackage[english]{babel}
\usepackage{multirow}
\usepackage{float}
\usepackage{url}
\usepackage{slashed}
\usepackage{xcolor} 
\usepackage[utf8]{inputenc}
\usepackage{comment}
\usepackage{physics}

\def\sfrac#1#2{{\textstyle{#1\over #2}}}
\newcommand{\be}{\begin{equation}}
\newcommand{\ee}{\end{equation}}
\newcommand{\ba}{\begin{array}}
\newcommand{\ea}{\end{array}}
\newcommand{\bea}{\begin{eqnarray}}
\newcommand{\eea}{\end{eqnarray}}

\newcommand{\nn}{\nonumber}

\begin{document}

\title{Asymmetric reheating from a symmetric inflationary potential}
\author{James M.\ Cline}
\email{jcline@physics.mcgill.ca}
\author{Jean-Samuel Roux}
\email{jsroux@hep.physics.mcgill.ca}
\address{McGill University, Department of Physics, 3600 University St., Montr\'eal, QC H3A2T8 Canada}

\begin{abstract}
We explore a model of two-field inflation with nonminimal kinetic terms in which two identical matter sectors decoupled from each other may reheat to different temperatures while preserving the symmetry of the Lagrangian. This scenario is motivated by mirror dark matter models in which the temperature of the mirror sector is constrained to be $T'\lesssim0.5 T$ by big bang nucleosynthesis and the cosmic microwave background. For a given class of nonminimal kinematic terms, we find that the symmetric field trajectory $X=Y$ is a repeller solution, such that any randomly-occurring asymmetry in the initial conditions is amplified by many orders of magnitude during inflation, far beyond  what canonical power-law models can achieve. Isocurvature fluctuations are strongly suppressed in this model, but a $\mathcal O(0.03$--$0.07$) tensor-to-scalar ratio could be observed in the near future.  The range of potential parameters compatible with
{\it Planck} constraints is shown to be much larger than 
in corresponding single-field models.  This occurs through a mechanism for lowering the spectral index that we
dub CTHC: curved trajectory at horizon crossing.
\end{abstract}
\maketitle

\section{Introduction}

Mirror models were perhaps the original paradigm for a hidden sector in particle physics \cite{Kobzarev:1966qya,Okun:2006eb,Foot:2014mia}.  A copy of the standard model (SM) field content and gauge group
is hypothesized, which in its simplest form is exact, so that the two sectors are
related by a discrete $Z_2$ mirror symmetry.  This form is subject to significant
cosmological constraints on the additional relativistic degrees of freedom---mirror photons $\gamma'$ and neutrinos $\nu'$---that would increase the Hubble expansion rate at early times, in contradiction to the successful predictions of big bang nucleosynthesis (BBN) and the cosmic microwave background (CMB).  To circumvent the bound on the effective number of neutrino species $N_{\rm eff}$, one must either break the mirror symmetry by making $\gamma'$
and $\nu'$ massive, or in the case of exact $Z_2$ symmetry, arrange for the initial conditions of the Universe to create a lower temperature $T'$
in the mirror sector than in the visible one, with  $x \equiv T'/T\lesssim 0.5$ \cite{Berezhiani:2000gw,Roux:2020wkp}.  


Similar frameworks like Twin Higgs models (see \cite{Curtin:2021alk} for a recent example) or parity solutions to the strong CP problem \cite{Craig:2020bnv} break the mirror symmetry at late times. However, if at least one mirror species like $\gamma'$ remains massless, those scenarios also require $x\lesssim 0.6$ to satisfy cosmological bounds on $N_{\rm eff}$. The breaking of the symmetry might facilitate entropy transfers that cool the mirror sector, but it is also possible that mirror and visible species decouple when the symmetry still holds, in which case the temperature hierarchy must originate from early universe dynamics, when the Lagrangian was still symmetric.

One may wonder how likely it is to realize perfect mirror
symmetry in a complete model including inflation, such that
the relative temperatures in the two sectors differ as 
required by the constraints. In Refs.\ \cite{Berezhiani:2000gw,Berezhiani:1995am,Berezinsky:1999az}, it was noted that asymmetric reheating would generically occur in models with two inflatons, one for each sector, due to differences in the initial conditions.  Here we revisit this idea, in the light of current CMB constraints from {\it Planck} \cite{Akrami:2018odb}.\footnote{An early proposal for getting asymmetric reheating was given in  ref.\ \cite{1985Natur.314..415K}, which proposed a `double-bubble inflation'  model where the  ordinary and mirror inflatons finish inflation by bubble nucleation at different (random) times. In this case the  first sector to undergo reheating gets exponentially redshifted until the  second field nucleates a bubble of true vacuum.  However this is in the context of ``old inflation'' driven by false vacua, which is untenable because the phase transitions never complete.}

We consider
two-field chaotic inflation with decoupled potentials of the form
\be \label{eq:pot2field}
V_{\rm tot} = V(X) + V(Y) =\lambda\frac{|X|^p}{m_P^{p-4}} +\lambda\frac{|Y|^p}{m_P^{p-4}}\,,
\ee
(where $m_P$ is the reduced Planck mass)
plus respective couplings of each field to its own sector's matter particles,  to accomplish reheating. In the case of a purely quadratic potential ($p=2$) the solutions are such that
the ratio of the two inflatons $Y/X$ remains constant
during inflation~\cite{Berezinsky:1999az}. Then the ratio of the reheating temperatures goes as 
$T'/T = (Y^2/X^2)^{1/4}$, and is thereby analytically determined
in terms of the initial conditions.  This example is now ruled out by \textit{Planck} data \cite{Akrami:2018odb},
which strongly disfavors chaotic inflation models that have convex potentials. 

 In the following, we study a model with noncanonical inflaton kinetic terms, proposed in Ref.~\cite{Roux:2020wkp}, that generates a temperature hierarchy by the spontaneous breaking of mirror symmetry
by the initial values of the inflatons.
At large field values, it is equivalent to Eq.\ (\ref{eq:pot2field}) 
with fractional values of $p$, and we therefore consider both kinds of
models. Such fractional power-law potentials have been proposed in the context of string theory \cite{Silverstein:2008sg,McAllister:2008hb,Dong:2010in,Gur-Ari:2013sba,McAllister:2014mpa,Marchesano:2014mla,Bielleman:2016grv,Landete:2017amp} or supergravity
\cite{Gao:2014fha}.  We will show, somewhat
surprisingly, that the initial values $X_i, Y_i$ cannot be too different from each other, while remaining consistent with {\it Planck} constraints on the CMB
observables; nevertheless, a small initial asymmetry is typically amplified
during inflation into a very large asymmetry in the final temperature ratio.
We will also show that the two-field inflationary scenarios lead to much better agreement with the CMB than their single-field counterparts, due to the effect of curvature in the inflaton trajectory in the $X$-$Y$ field space, at the time of
horizon crossing.  We abbreviate this effect by ``CTHC.''  It was previously observed in the context of multifield inflation with fractional power law potentials in Ref.\ \cite{Wenren:2014cga},
with emphasis on many inflatons having random potential parameters.

In Section \ref{sectII} we introduce the noncanonical two-field models, and the
corresponding canonical models that are equivalent at large field values.  The
numerical techniques used are described there, along with analytical approximations that explain the qualitative behavior of the exact solutions.  We illustrate our results for three benchmark parameter choices in Section \ref{sec:benchmark}.
This is followed in Section \ref{sectIV} by a description of a Monte Carlo search of the full parameter space for models that are consistent with all constraints.
We give conclusions in Section \ref{sectV}.

\section{Noncanonical two-field inflation}
\label{sectII}

 {\it Planck} data favor concave inflaton potentials,
whereas convex ones occur more generically.  A popular solution to this problem is through nonminimal coupling to
gravity (see {\it e.g.,} Ref.\ \cite{Linde:2011nh}), which in our framework would take the form
\be
    {\cal L}\ni\sfrac12 m_P^2\, \zeta\, R\, (X^2 + Y^2)\,,
\ee
where $R$ is the Ricci curvature.  Transforming to the Einstein frame, the inflaton potential gets rescaled by
$V(X,Y) \to V(X,Y)/\Omega^4$ where $\Omega^2 = 1 + \zeta(X^2+Y^2)$, so that $V$ becomes concave at large field values.  

However this simple device, while reconciling chaotic inflation potentials with CMB data, introduces a potentially strong coupling between the two fields, through the factor $\Omega^{-4}$.  By numerical investigation we found that whenever $\zeta$ is large enough to resolve the
tensions with CMB observations, it also causes the trajectories to align in the inflationary attractor solutions, $X\cong Y$, such that the effect of random initial differences $X_i \neq Y_i$ gets erased rather than enhanced during inflation, leading to nearly equal reheat temperatures in the two sectors.  This is not compatible with the goal of the present work.

\subsection{Noncanonical models}
Another way of reconciling chaotic inflation potentials with
CMB constraints is to use nonminimal kinetic terms, for instance
\cite{Lee:2014spa}, 
\begin{align} 
{\cal L} &= \frac{1}{2}\left(1+ f {\frac{X^n}{m_P^n} }\right) (\partial X)^2 + 
\frac{1}{2}\left(1+ f {\frac{Y^n}{m_P^n}}\right) (\partial Y)^2 \nn \\ 
& + \frac{1}{2}\, m^2 \left(X^2+Y^2\right)\, . \label{eq:nmk}
\end{align}
For $X,Y\gg m_P$, the canonically normalized fields are $U \sim X^{1+n/2}$,
$W\sim Y^{1+n/2}$, so that the potential
becomes proportional to $(U^{4/(n+2)} + W^{4/(n+2)})$. Here we will take the noncanonical fields to have a quadratic potential, as the simplest and most generic example.

 The model of Eq.~(\ref{eq:nmk}) essentially coincides with the power-law potential of Eq.~(\ref{eq:pot2field}) at large field values, with $p=4/(2+n)$. The two scenarios only differ significantly near the end of inflation when the fields are close to the minimum of the potential, and most results from Ref. \cite{Berezinsky:1999az} also apply to our noncanonical scenario.

In the slow roll approximation, the equations of motion
$3H\dot X = -\partial V/\partial X$ and $3H\dot Y = -\partial V/\partial Y$ for the potential of Eq.\ (\ref{eq:pot2field}) can be integrated to 
show that \cite{Berezinsky:1999az}
\begin{align}
\nn X/Y &= {\rm const}, \quad  p=2;\\
    X^{2-p} - Y^{2-p} &= {\rm const},\quad  p\neq 2.\label{eq:srtraj}
\end{align} 
Unlike the quadratic scenario, for $p\neq 2$ the inflaton trajectory is not a straight line in the $(X,Y)$ field space, unless
the initial values $X_i$ and $Y_i$ are taken to be equal (or if one of them remains
zero throughout inflation).  If $p>2$, the two fields become synchronized before the end of inflation, since $X$ and $Y$ are both decreasing, and $X\approx Y$ when reheating takes place. This generally leads to both sectors reaching thermal equilibrium by the time of BBN, a possibility that is ruled out by the current bounds on $N_{\rm eff}$.

\begin{figure}[t!]
\begin{center}
 \centerline{\includegraphics[scale=0.375]{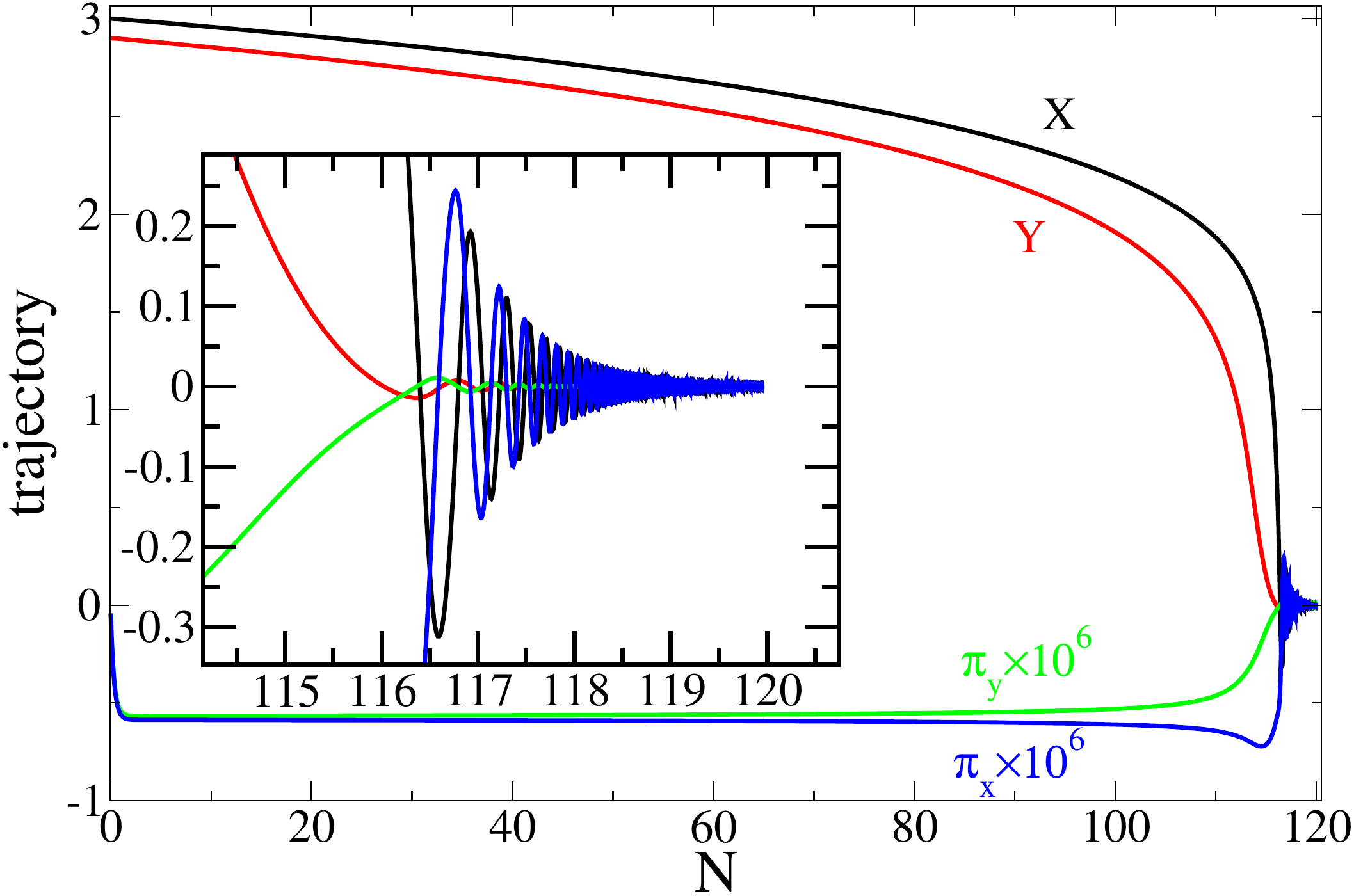}}
\caption{Illustrative trajectory for inflatons starting from nearly
symmetric initial conditions in the model of Eq.\ (\ref{eq:nmk}), but achieving asymmetric reheating due to inflation ending earlier for the mirror inflaton $Y$.  The inset highlights the end of inflation.}
\label{fig:inftraj}
\end{center} 
\end{figure}

The case $p<2$ is the most appealing for our purposes, since the fields are driven away from the symmetric trajectory $X=Y$. Then any initial randomly-occurring asymmetry is amplified by inflation, possibly leading to a difference of many orders of magnitude between the respective reheat temperatures. Moreover, 
potentials with $p\sim 2/3$ \cite{McAllister:2008hb,Silverstein:2008sg}
are generally consistent with \textit{Planck} constraints on the tensor-to-scalar ratio $r$ and spectral index $n_s$ \cite{Akrami:2018odb}. However, in single-field inflation this agreement is marginal at best and is limited to a small range of powers $p$, as flatter potentials lead to to too-large predictions for $n_s$,
and steeper ones give too-large $r$ \cite{Planck:2013jfk}. We will show that in the two-field models the consistency is improved, for both the canonical and noncanonical implementations, and for a larger range of $p$ relative to single-field models.

 In Fig.\ \ref{fig:inftraj} we show an example of field trajectories for the noncanonical Lagrangian of Eq.~(\ref{eq:nmk}) with an effective $p=2/3$ power-law potential ($n=4$) and  parameter values\footnote{Throughout this work we fix the inflaton mass to $m=10^{-6}$ in the noncanonical model (no explicit mass term is present in the canonical version). Its value does not impact the inflationary dynamics during slow roll, but in a complete model it would set the scale of reheating, the details of which are not considered here.}
\be
 f = 1,\ X_i = 3,\ Y_i = 2.9\,,
\ee
where here and henceforth we work in Planck units with $m_P=1$, and assume
vanishing initial velocities.  Even though the initial values in this example
differ only by 3\%, slow roll ends somewhat earlier in the $Y$ field,
and its amplitude gets damped by a factor of $\sim 25$ relative to
that of $X$, leading to a temperature ratio of $x\sim 0.2$, in agreement with the cosmological bound $x\lesssim0.5$.  Although we do not discuss the detailed mechanism of reheating in this work, the assumptions of unbroken mirror symmetry
and complete conversion of the inflaton energies into radiation in the two respective sectors allows us to estimate the temperature ratio as
\be
x\equiv{T'\over T} = \left(\rho'\over\rho\right)^{1/4}\,,
\label{Trheq}
\ee
by evaluating the energy densities at the end of inflation.

\subsection{Numerical solution}

To solve for the inflationary dynamics, we use the first order 
equations of motion for numerical integration, in terms of the canonical momenta
$\pi_x = \partial{\cal L}/\partial\dot X = F(X)\dot X$, $\pi_y = \partial{\cal L}/\partial\dot Y = F(Y)\dot Y$, with $F(X) \equiv 1 + f X^n$ for noncanonical models.  It is convenient to take the number of $e$-foldings
$N$ as the independent variable, defined through $dN=H dt$ in terms of the
Hubble parameter $H$.  Denoting $d/dN$ by primes,  the system of equations is
\bea
    X' &=& {\pi_x\over F(X) H}\,, \nn\\
    \pi_x' &=& -3\pi_x  + {1\over H}\left({F'(X)\over 2}\left(\pi_x\over F(X)\right)^2 - m^2 X\right)\,,\nn\\
        Y' &=& {\pi_y\over F(Y) H}\,, \nn\\
    \pi_y' &=& -3\pi_y  + {1\over H}\left({F'(Y)\over 2}\left(\pi_y\over F(Y)\right)^2- m^2 Y\right)\,,\nn\\
    H &=& {1\over \sqrt{6}}\left({\pi_x^2\over F(X)} + {\pi_y^2\over F(Y)}
    + m^2(X^2+Y^2)\right)^{1/2}\,.
\eea

Since we assume the two fields are decoupled from each other, each inflaton  decays into its own matter sector during reheating. The resulting temperature ratio can therefore be estimated as in Eq.\ (\ref{Trheq}),
\be 
x=  \left( \frac{\frac{1}{2}F(Y) (\partial Y)^2 + V(Y)}{\frac{1}{2}F(X) (\partial X)^2 +V(X)} \right)^{1/4}\,,
\ee 
evaluated at the end of inflation.
Without loss of generality, we will assume $Y<X$ initially, so that $x<1$
and $X$ denotes the visible sector inflaton.

\subsection{Slow-roll parameters}

To compute the slow-roll parameters and  inflationary observables it is convenient to work in the canonical basis, with fields denoted by $(U,W)$. Since the two kinetic terms in Eq.~(\ref{eq:nmk}) are decoupled from each other, the Jacobian matrix $Z$ is diagonal,
\be \label{jacobian}
\begin{pmatrix}
\dot X \\ \dot Y 
\end{pmatrix} = 
\begin{pmatrix}
F(X)^{-1/2} & 0\\
0 & F(Y)^{-1/2}
\end{pmatrix} 
\begin{pmatrix}
\dot U\\
\dot W
\end{pmatrix}\equiv Z \begin{pmatrix}
\dot U\\ \dot W
\end{pmatrix}.
\ee 

The slow-roll parameters computed in the canonical basis (indices $m,n$) are related to derivatives with respect to fields in the the original basis (indices $i,j$) by
\begin{align}
\epsilon_m &= \frac{(Z_{im} \partial_i V)^2}{2V^2} \,, \nn\\
\eta_{mn} & = Z_{im}Z_{jn} \frac{\partial_{ij}V}{V} + Z_{im}\partial_{i}Z_{jn} \frac{\partial_{j}V}{V}.
\end{align}
For numerical purposes, we modify these definitions by replacing $V\to\rho = 3 H^2$ in the denominators.  During inflation, this makes a negligible difference, whereas at the end of inflation while the inflaton is oscillating around its minimum, it avoids the artificial singularities that result from $V$ passing through zero.

The formalism for computing the spectral index $n_s$ of primordial adiabatic perturbations and the tensor-to-scalar ratio $r$ in two-field models was developed
in Refs.\ \cite{Gordon:2000hv,Byrnes:2006fr}.  One first introduces the adiabatic/entropy basis $(\sigma,s)$, defined by
\bea
	d\sigma &=& c_\alpha dU + s_\alpha dW\,,\nn\\
	ds &=& -s_\alpha dU + c_\alpha dW\,,
\label{dseq}
\eea
where $\alpha = \tan^{-1}(\dot W/\dot U)$ is the instantaneous slope of the field trajectory in the $U$-$W$ plane. The slow-roll parameters in the $(\sigma,s)$ basis are computed using 
\cite{Gordon:2000hv}
\bea
	\epsilon_\sigma &=& (c_\alpha \partial_U V + s_\alpha 
	\partial_W V)^2/
(2 V^2)\nn\\
	\epsilon_s &\simeq& 0\nn\\
	\eta_{\sigma\sigma} &=& c_\alpha^2 \eta_{UU} 
	+ s_\alpha^2 \eta_{WW}\nn\\
	\eta_{ss} &=& s_\alpha^2 \eta_{UU} 
	+ c_\alpha^2 \eta_{WW}\nn\\
	\eta_{\sigma s} &=& c_\alpha s_\alpha (\eta_{WW} - \eta_{UU}). \label{eq:srparams}
\eea
and the fact that $\eta_{UW}=0$ because $Z$ is diagonal [\emph{cf.}\ Eq.~(\ref{jacobian})].  The derivatives of the potential in the $(U,W)$ basis are computed using the Jacobian matrix, $\partial_m V = Z_{im} \partial_i V$.
Then to leading order in the slow-roll expansion,
the scalar spectral index and tensor-to-scalar ratio are \cite{Byrnes:2006fr}
\bea
\!\!\!\!\!\!\!\!\!n_s -1 &=&-(6-4 c_\Delta^2)\epsilon_\sigma \!+\! 2 s_\Delta^2
\eta_{\sigma\sigma}\! +\! 4 s_\Delta c_\Delta \eta_{\sigma s}\!+\!2 c_\Delta^2 \eta_{ss}\,,\nn\\
	r &=& 16\epsilon_\sigma\,, \label{eq:nsr}
\eea
where $c_\Delta = -2{\cal C}\,\eta_{\sigma s}$, $s_\Delta =
+\sqrt{1-c_\Delta^2}$, ${\cal C} = 2-\ln 2-\gamma\simeq 0.73$
($\gamma$ is the Euler constant). 

For the interpretation of the following results, it is useful to derive approximate expressions for $n_s$ and $r$ in terms of the
background field amplitude and the canonical field ratio, defined
respectively as 
\be
    \sigma\equiv(U^2+W^2)^{1/2},\quad \xi\equiv {W\over U}\,.
\ee
This will help to elucidate how the two-field model improves over single-field inflation. In the large field limit, $\xi\cong \dot{W}/\dot{U}$, which is equivalent to approximating the instantaneous field velocity as being radial in the $U$-$W$ plane. This implies that $\cos\alpha\approx U/\sigma$ and $\sin\alpha\approx W/\sigma$ in Eqs.~(\ref{dseq}-\ref{eq:srparams}). One can further approximate $\eta_{\sigma s}\ll1$ so that $\cos^2\Delta\sim0$ in Eq.~(\ref{eq:nsr}). Hence we obtain
\begin{align} \nn
n_s -1& \approx  - \frac{p(p+2)}{\sigma_*^2} - \frac{8\mathcal{C}p^2(p-1)^2}{\sigma_*^4}\left(\frac{\xi_*^p-\xi_*^2}{\xi_*(1+\xi_*^p)}\right)^2,\\
r &\approx \frac{8 p^2}{\sigma_*^2},
\label{eq:nsapprox}
\end{align} 
where $\sigma_*$ and $\xi_*$ are evaluated at horizon crossing, $N_*\approx \sigma_*^2/2p$.  The extra term in $n_s-1$, which is negative and arises from the
term $4 s_\Delta c_\Delta \eta_{\sigma s}$ in Eq.\ (\ref{eq:nsapprox})), explains why the spectral index comes into better agreement with {\it Planck} data than in single-field chaotic inflation.  This term is explicitly related to the curvature of the potential and ensuing slow-roll trajectory, inspiring our CTHC (curved trajectory at horizon crossing) acronym.
The expression for $r$ is identical to that for single-field inflation \cite{Planck:2013jfk}.\footnote{In the numerical results, $r$ also depends weakly on $\xi_*$, but this dependence does not appear in Eq.~(\ref{eq:nsapprox})  at the level of approximations we have used; to leading order $r$ is determined by $N_*$.}

These are rough estimates, since Eq.~(\ref{eq:srtraj}) indicates $\dot{W}/\dot{U}=\xi^{p-1}$ during slow roll, in contrast to the approximation we have used. For small values of  $\xi_*\lesssim 0.1$, the mirror inflaton is no longer slowly rolling and then
higher order contributions to 
Eq.~(\ref{eq:nsr}) become important.  One can clearly see this breakdown in the approximations  in the limit $\xi_*\to 0$, where the predictions for single-field
inflation would be recovered in an exact expression.  Nevertheless, 
 Eq.~(\ref{eq:nsapprox}) accurately describes the correlations between $\sigma_*$, $\xi_*$ and the cosmological observables when the two scalar fields are both slowly rolling during horizon exit, which is satisfied for $\xi_*$ not too much less than unity.

\begin{table*}[t]
\centering
\begin{tabular}{|c|c||c|c|c|c|c|c|c||c|c|c|c|}
\hline
&& \multicolumn{7}{c||}{Noncanonical} & \multicolumn{4}{c|}{Canonical} \\
\hline
  $p$ & $x_i$ & $F(X)$ & $X_i$ & $Y_i$ & $f$ & $N_{\rm tot}$ & $x_{f}$ & $\beta$ & $X_i$ & $Y_i$ & $N_{\rm tot}$ & $x_f$\\
\hline
$2/3$ & $0.91$ &  $1+fX^4$  & $3.0$ & $2.48$ & $0.78$ & $68.1$  &  $4.2\times 10^{-3}$ & $5.8\times 10 ^{-15}$ & $8.4$ & $4.77$ & $71.6$ & $2.8\times 10^{-4}$ \\ 
$1/2$ & $0.96$ & $1+fX^6$ & $2.0$ & $1.84$ & $4.5$ & $112.6$ & $5.2\times 10^{-7}$ & $1.6\times 10^{-20}$ & $8.6$ & $6.16$ & $113.8$ & $1.2\times 10^{-3}$ \\ 
$1/3$ & $0.987$ &  $1+fX^{10}$  & $1.6$ & $1.56$ & $12$ & $247.9$ &  $2.1\times 10^{-8}$ & $1.8\times 10^{-32}$& $9.7$ & $8.33$ & $247.8$ & $4.5\times 10^{-3}$  \\ 
\hline   
\end{tabular}
\caption{Parameters and initial values for three noncanonical benchmark models and their corresponding canonical power-law scenarios, $V\sim |X|^p+|Y|^p$, where $p=4/(n+2)$. Parameters include the
total number of $e$-foldings of inflation $N_{\rm tot}$, the initial and final values of the temperature ratio $x$, and the amplitude of the isocurvature power spectrum on large scales $\beta$.
\label{tab1}}
\end{table*}

\subsection{Isocurvature fluctuations}
Two-field inflation models have the potential of generating isocurvature contributions to the power spectrum $\mathcal{P}_S(k)$, which are constrained by {\it Planck}. Thus, we compute the evolution of entropy perturbations to estimate their amplitude after inflation. 

One can relate the adiabatic and entropy fluctuations at the end of inflation to their values at horizon crossing using a matrix of transfer functions \cite{Byrnes:2006fr,Amendola:2001ni,Wands:2007bd},
\be \label{transfer}
\begin{pmatrix}
\mathcal{R}\\ \mathcal{S}
\end{pmatrix} = \begin{pmatrix}
T_{RR} & T_{RS} \\ T_{SR} & T_{SS}
\end{pmatrix} \begin{pmatrix}
\mathcal{R}_* \\ \mathcal{S}_*
\end{pmatrix},
\ee 
where the dimensionless adiabatic and isocurvature fluctuations are given by $\mathcal{R} = H \delta \sigma /\dot{\sigma}$ and $\mathcal{S} =H \delta s/\dot{\sigma}$  in the spatially flat gauge, and the star subscript indicates the time of horizon crossing. The matrix elements in the upper and lower rows are obtained by solving the perturbed equations of motion with initial conditions $(\mathcal{R}_*, \mathcal{S}_*)=(1,0)$ and $(0,1)$ respectively. 

To compute the transfer functions of Eq.~(\ref{transfer}), it is more convenient to instead consider the evolution of the canonical fields' perturbations $\delta U,\delta W$ by making use of Eq.~(\ref{dseq}). Their equations of motion are \cite{Cline:2019fxx}
\bea
\nn \delta U'' &= -C_1\, \delta U' - 3\eta_{UU}\,\delta U + U'\, \delta C\\ 
\nn & + (U'^2)'\,\delta U + (U'W')'\,\delta W \\
\nn \delta W'' &= -C_1\, \delta W' - 3\eta_{WW}\,\delta W + W'\, \delta C\\ 
& + (W'^2)'\,\delta W + (U'W')'\,\delta U. 
\eea
Primes denote $d/dN$, $C_1=3+H'/H$ and $\delta C=C_1(U'\,\delta U+W'\, \delta W)$.

The amplitude of the isocurvature power spectrum is characterized by the scale-dependent primordial isocurvature fraction, $\beta(k) = \mathcal{P}_S(k)/[\mathcal{P}_S(k)+\mathcal{P}_R(k)]$, where $\mathcal{P}_R(k)$ denotes the adiabatic power spectrum. On large scales, the leading contribution is \cite{Byrnes:2006fr} 
\be 
\beta \cong \frac{T_{SS}^2}{1+T_{SS}^2+T_{RS}^2}\lesssim\mathcal{O}(10^{-3}\text{--}10^{-2})\,.
\ee 
The experimental upper limit from {\it Planck} \cite{Akrami:2018odb} depends on the assumptions of the fit and which datasets are used.

\section{Benchmark models}
\label{sec:benchmark}

\begin{figure}[t]
\begin{center}
 \centerline{\includegraphics[width=\linewidth]{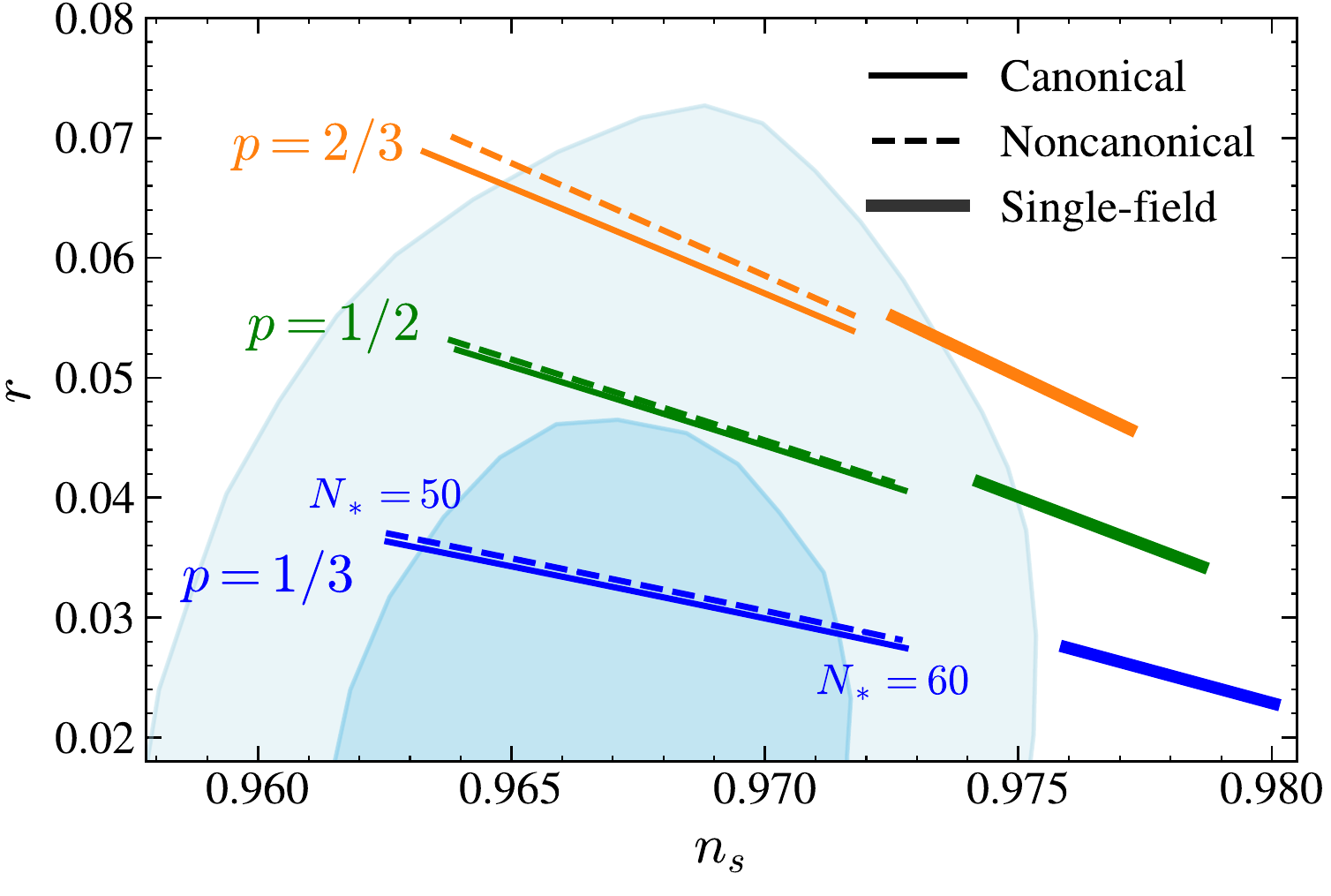}}
\caption{Spectral index and tensor-to-scalar ratio for the models of Table \ref{tab1} for $N_*\in[50,60]$. The shaded regions show the 68 and 95\% C.L. limits set by {\it Planck}. For noncanonical models, the effective power $p$ corresponds to $n=4/p-2$ in Eq.~(\ref{eq:nmk}). The thick lines to the right show the predictions in corresponding single-field chaotic inflation with the same fractional power-law potential.}
\label{fig:nsr}
\end{center} 
\end{figure}

We start by considering three sets of benchmark parameters that
illustrate the possibilities for phenomenologically successful inflation with a large reduction of temperature in the mirror sector.  The results are presented for both the noncanonical models and their corresponding canonical versions 
in Table~\ref{tab1} and in Fig.~\ref{fig:nsr}.  Initial conditions
were chosen such that the initial temperature ratios $x_i=(\rho'_i/\rho_i)^{1/4} \sim 0.91-0.99$ were the same in both types of models. It can be seen in the figure that corresponding models lead to similar predictions for $n_s$ and $r$ (solid versus dashed lines).

The canonically normalized models with fractional power potentials
are numerically challenging to evolve at late times while the
fields are oscillating. During this regime, the fields undergo damped anharmonic motion and their energy density redshifts as \cite{Turner:1983he}
\be \label{eq:rhoscaling}
\rho\sim a^{-6p/(2+p)}, 
\ee 
where $a$ is the scale factor. As the amplitudes decrease, their frequency increases rapidly, impeding accurate numerical evolution. Hence for the canonical models we stop following the evolution of the mirror inflaton $Y$ once it reaches its minimum and instead extrapolate its energy density from the onset of oscillations to later times using Eq.~(\ref{eq:rhoscaling}).

The amplitude of the isocurvature power spectrum $\beta$ in noncanonical models was estimated by assuming a number of $e$-foldings at horizon crossing $N_*=55$. In every case, the predicted amplitude is much smaller than current bounds set by \textit{Planck}.\footnote{A similar conclusion applies for the model of Ref.\ \cite{Cline:2019fxx}, which was not recognized in that work.}   This can be understood from the field trajectory of Fig.~\ref{fig:inftraj}: since $Y$ reaches its minimum before the end of inflation, its perturbations are strongly diluted by the continuing exponential expansion of the universe. Although we could not determine  $\beta$ very precisely in canonical power-law models, one expects that since the general dynamics are similar to noncanonical models, they should likewise lead to insignificant isocurvature perturbations.

All of the benchmark noncanonical models and their canonical counterparts lead to temperature ratios that are safely below the cosmological bound $x_f<0.5$, despite the initial ratio being close to $1$ at the beginning of inflation. However, for $p<2/3$, noncanonical models lead to values of $x_f$ drastically lower than their canonical counterparts, $x\ll10^{-3}$, a trend we will confirm below in a Monte Carlo analysis. In this limit the mirror sector would be almost unpopulated after reheating, an assumption that was made in \textit{e.g.} Refs.~\cite{Foot:2014mia,Foot:2011ve}. To our knowledge, the scenario presented here is the first one that can consistently predict such a cold mirror sector, consistent with CMB constraints, without requiring significant fine tuning in the initial conditions. 

\begin{figure*}[t!]
\begin{center}
\centering
 \includegraphics[width=0.49\linewidth]{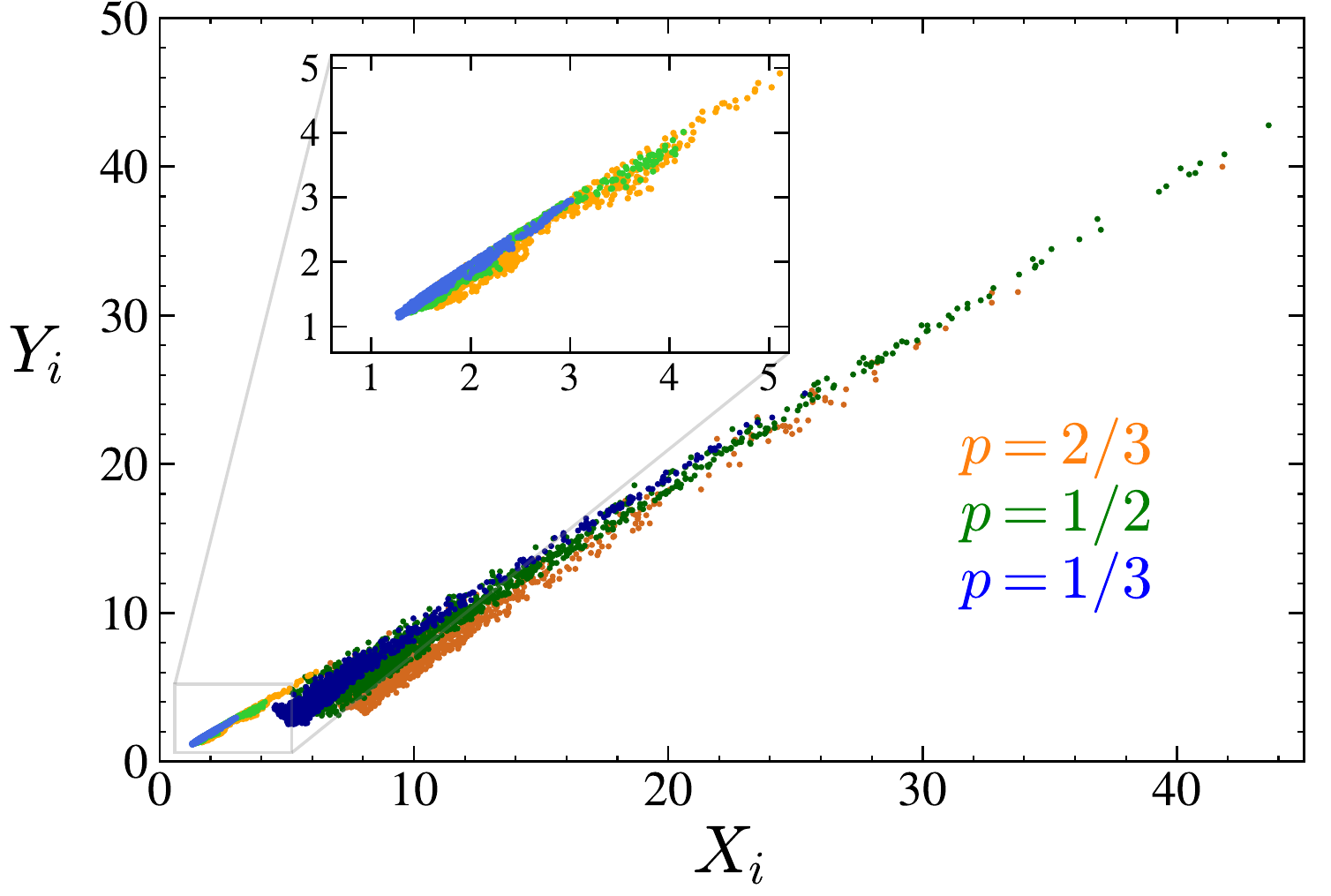}\hfill
 \includegraphics[width=0.49\linewidth]{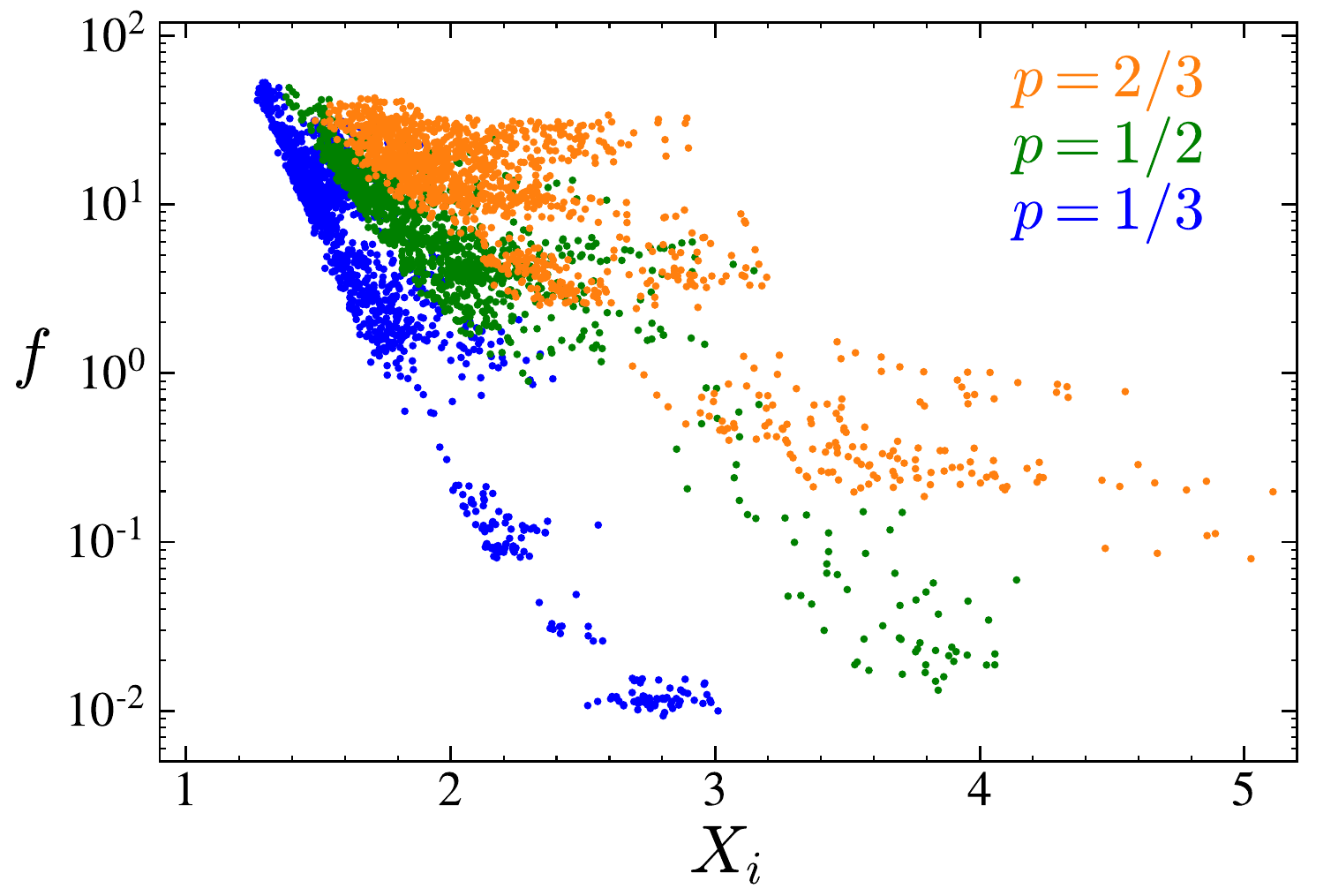}
\caption{Left: correlation of initial conditions in the MCMC chains. Dark (light) colors correspond to canonical (noncanonical) models. The inset shows the noncanonical models only. Right: correlation of $f$ versus $X_i$ for noncanonical models.}
\label{fig:mcmcparam}
\end{center} 
\end{figure*}

\subsection{Comparison with single-field power law inflation}

For a fixed initial temperature ratio $x_i$, the two types of models predict similar values for $n_s$ and $r$ if initial conditions are chosen such that the total duration of inflation $N_{\rm tot}$ is approximately the same, as can be seen in Table \ref{tab1}
and in Fig.~\ref{fig:nsr}.  This follows from the fact that they
coincide in the large field limit, and differ only in their behavior near the end of inflation.  As predicted by Eq.~(\ref{eq:nsapprox}), lower values of $p$ lead to lower predictions for the tensor-to-scalar ratio $r$, which tends to zero in the limit $p\to 0$ ($n\to\infty$). We can extrapolate our results to predict that a value of $r$ below $0.01$ would require a power-law potential with $p\lesssim0.15$, that is, $n\gtrsim 25$.

It is striking that the two-field inflationary scenarios are generally in much better agreement with \textit{Planck} data than
their single-field counterparts, represented by thick lines in Fig.~\ref{fig:nsr}. In single-field models, a flatter potential (lower $p$) decreases $r$ but also makes $n_s$ closer to 1, such that $p=2/3$ and $p=1/2$ models are marginally consistent with \textit{Planck} while $p=1/3$ is disfavored \cite{Planck:2013jfk}. 

With two scalar fields rolling together, the Hubble parameter is increased by a factor of $\sqrt{2}$, effectively making each field roll more slowly at a given amplitude.\footnote{`Assisted inflation' and `N-flation' models are based on this idea \cite{Liddle:1998jc, Dimopoulos:2005ac}. With $N$ independent scalar fields, $H$ is increased by $\sqrt{N}$, and the dynamics mimics that of a field at a higher scale.} However this does not explain the better agreement of our models, since this effect preserves the dependences of $n_s$ and $r$ on $N_*$.  In fact, if we set initial conditions $X_i=Y_i$,  the failure of the single-field models persists, as can be seen by letting $\xi_*=1$ in Eq.~(\ref{eq:nsapprox}).

Instead, the values of $n_s$ and $r$ in the two-field models 
depend on the difference in times for the end of inflation in the two sectors.
As the mirror inflaton approaches its potential minimum, the field trajectory deviates from a straight line, and the extra curvature of the potential contributes to making the spectral index lower than what single-field inflation can achieve. This is the CTHC mechanism described by Eq.~(\ref{eq:nsapprox}): since the field amplitude at horizon exit $\sigma_*$ is set by $N_*$, the main quantity that impacts $n_s$ is the canonical field ratio $\xi_*=W_*/U_*$. $\xi$ is strictly decreasing during inflation, and Eq.~(\ref{eq:nsapprox}) is a monotonically increasing function of $\xi$ in the interval $[0,1]$. This implies that $\xi_*$ must fall within a precise window of values at $N_*$ to give optimal agreement with CMB observations [see Eq.\ (\ref{optrange})].  Moreover, if the field  ratio at the beginning of inflation $\xi_i$ is too small, that is, if the initial asymmetry between the fields is too big, the value of $n_s$ will be too small. In other words, the agreement between cosmological observations and the predictions of our model imposes a lower bound on the field ratio at the beginning of inflation (assuming that $Y_i < X_i$). We will quantify this bound and the optimal range of $\xi_*$ at horizon crossing in the next section.

\begin{figure*}[t!]
    \centering
    \includegraphics[width=0.49\linewidth]{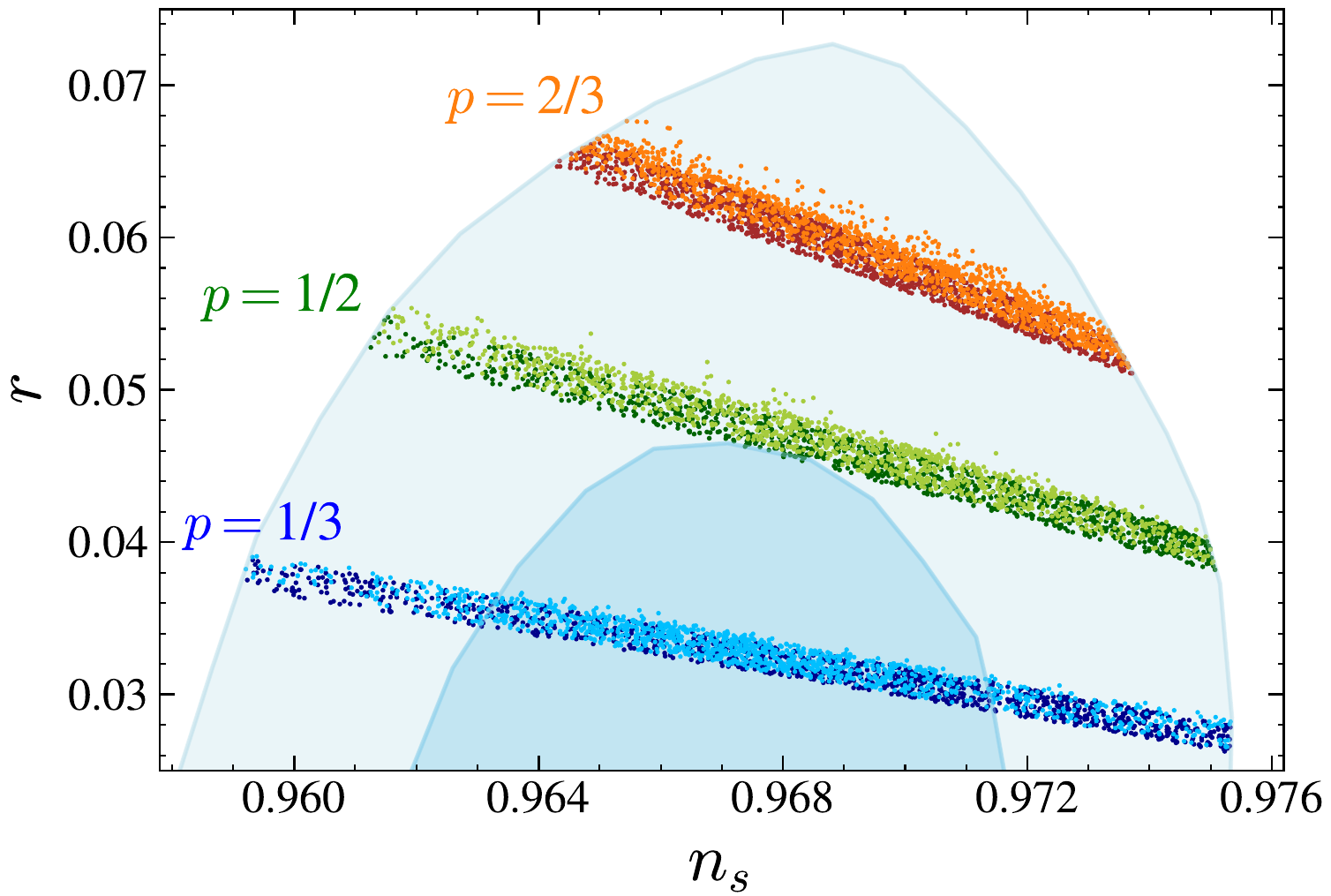} \hfill
    \includegraphics[width=0.49\linewidth]{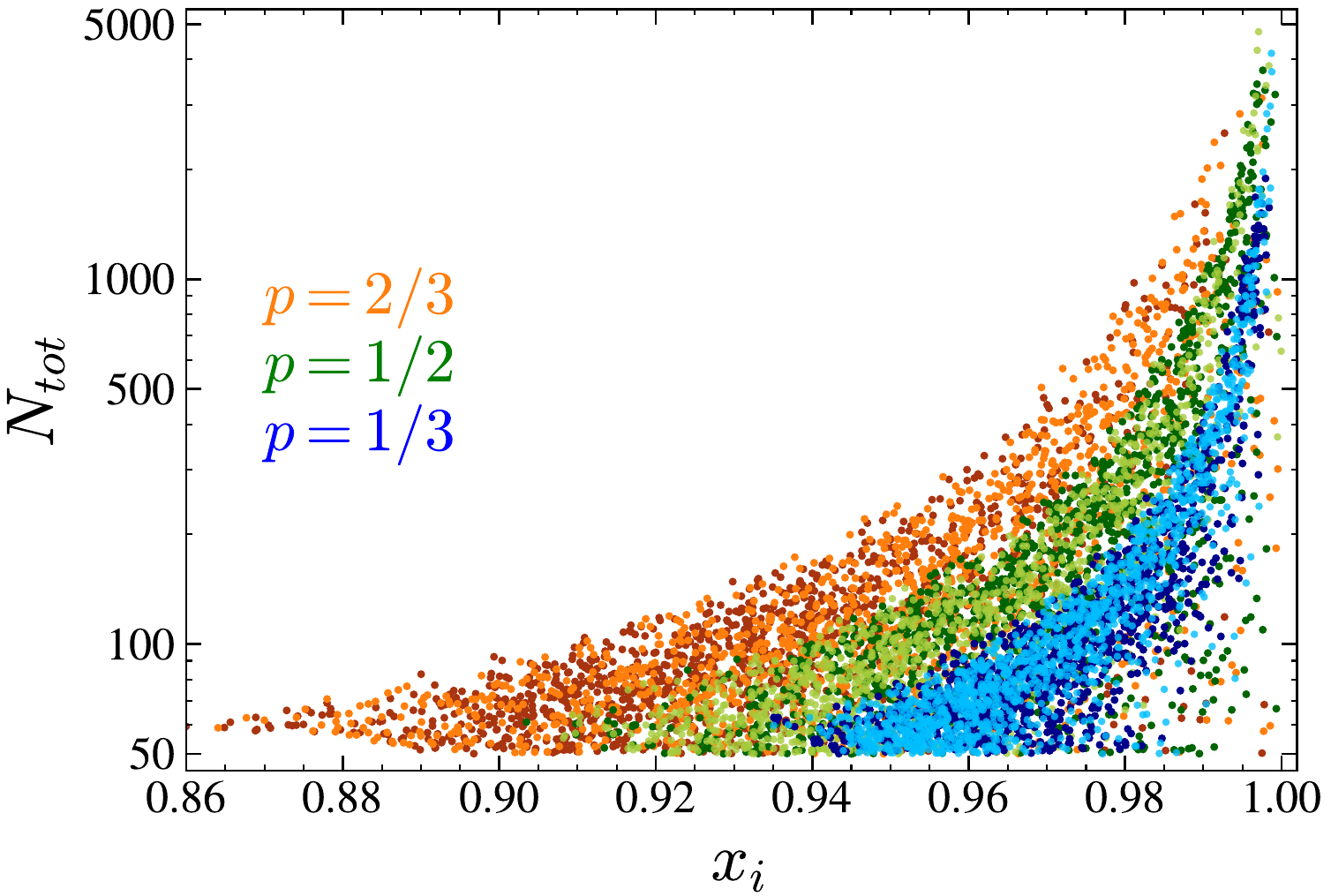}
    \caption{Scatter plots from the Markov Chain Monte Carlo (MCMC) analysis. Left: correlation between $n_s$ and $r$. Right: correlation of the initial temperature asymmetry $x_i=T'_i/T_i$ with the total number of $e$-foldings during inflation $N_{tot}$. Dark (light) colors correspond to canonical (noncanonical) models, but there is no significant distinction between the two types of scenarios within these plots. All models are within the 95\% C.L.\ limits of \textit{Planck}.}
    \label{fig:scatter1}
\end{figure*}

However, if $\xi_*\ll1$ at horizon crossing, the approximate results of Eq.~(\ref{eq:nsapprox}) are invalid and we instead recover the single-field limit (\textit{cf} the thick lines in Fig.~\ref{fig:nsr}), since the
mirror field makes a negligible contribution to inflation. Here the final temperature ratio can be arbitrarily small and isocurvature perturbations are essentially nonexistent, but the spectral index and the tensor-to-scalar ratio are at best marginally consistent with CMB constraints. In the remainder of this work we will focus on the two-field inflationary models, where the mirror inflaton is still slowly rolling at horizon crossing.

Because the field ratio $\xi$ is closely related to the temperature ratio $x=T'/T\approx(W/U)^{p/4}$, Eq.~(\ref{eq:nsapprox}) further implies a correlation between $n_s$ and $x_f$. This will allow us to identify ranges of $x_f$ that minimize the $\chi^2$ of observed versus predicted $n_s$ and $r$ values in the following section, that is, ranges of $x_f$ that are most consistent
with CMB observations.

\begin{figure*}[t!]
    \centering
    \includegraphics[width=0.49\linewidth]{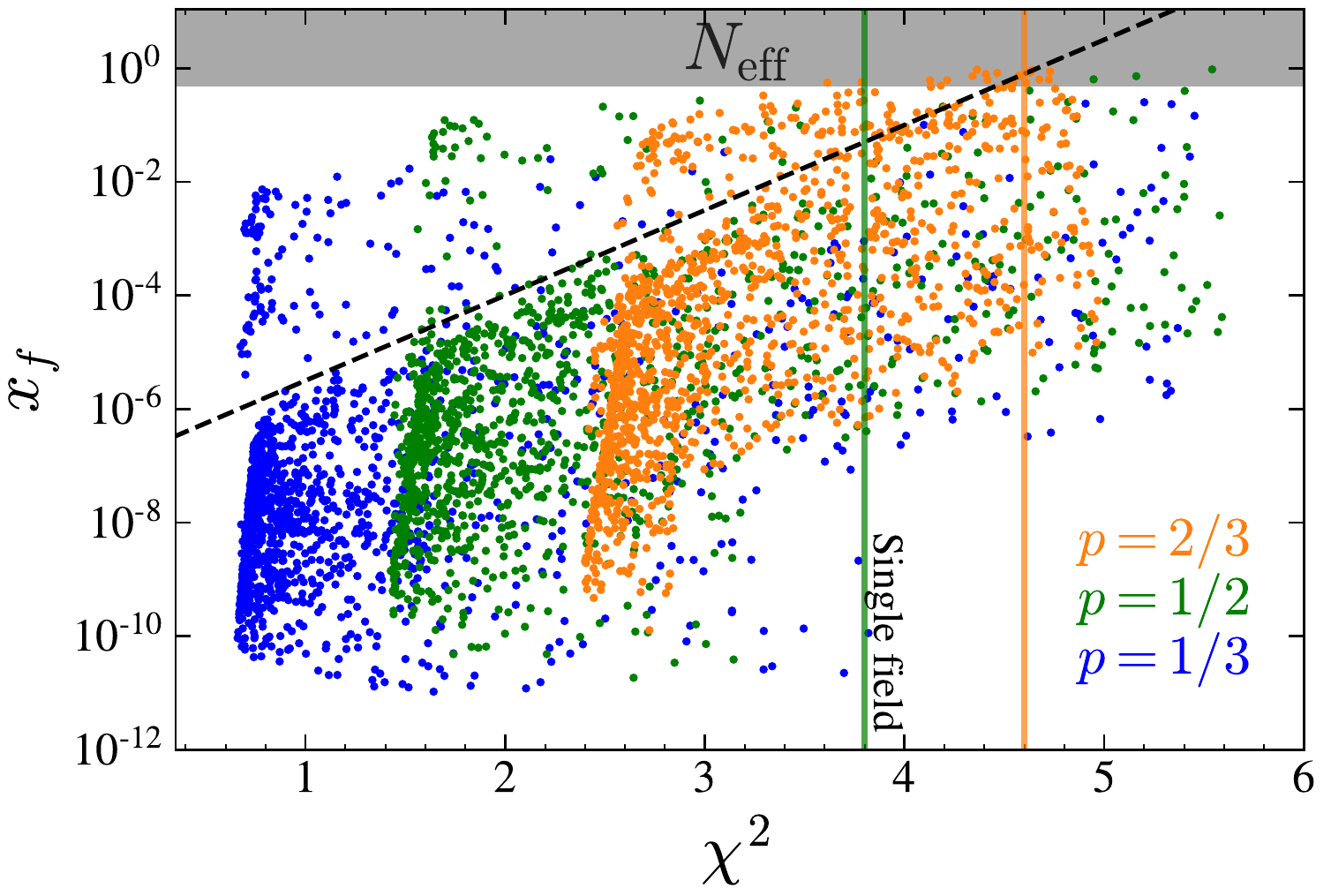} \hfill
    \includegraphics[width=0.49\linewidth]{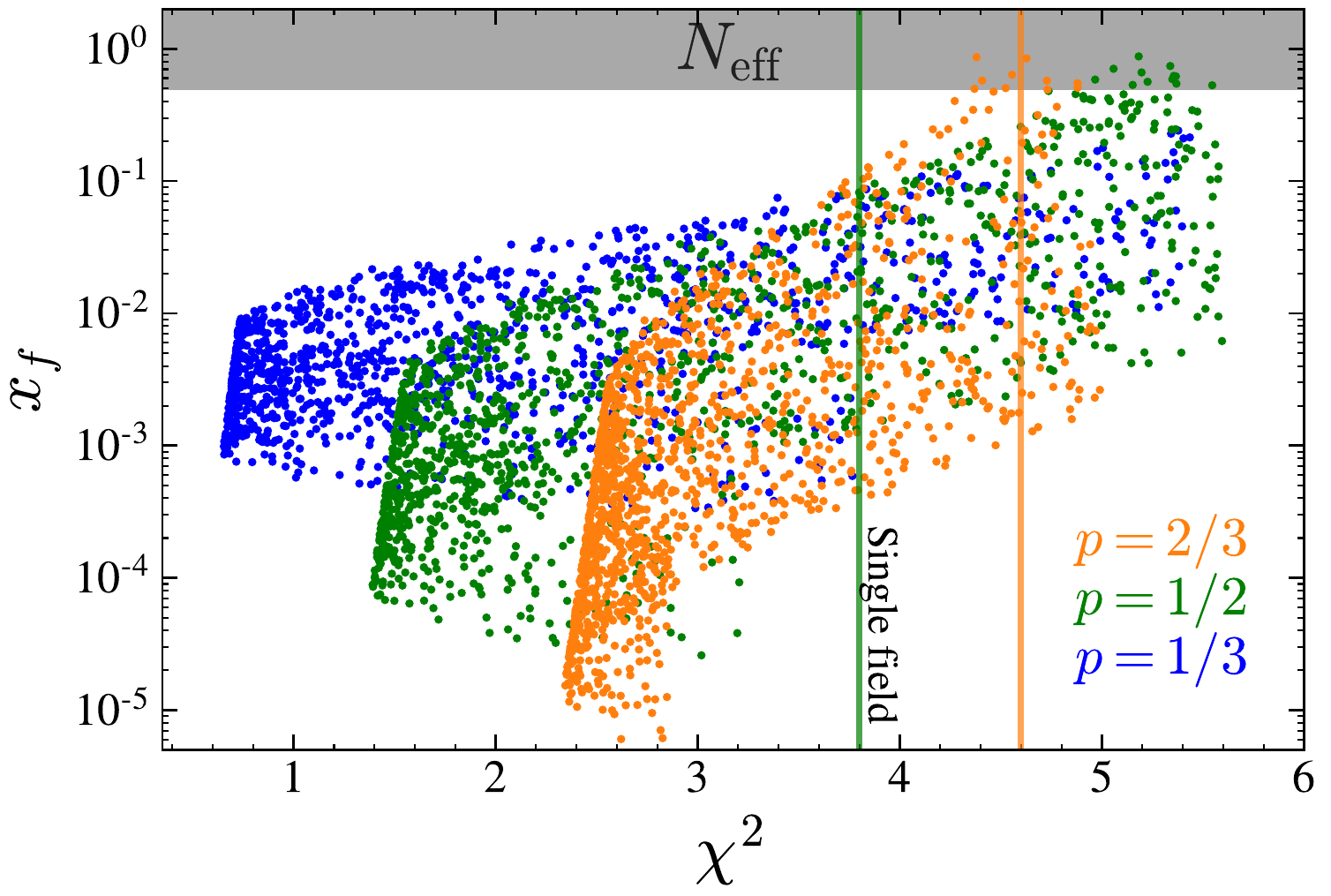}
    \caption{Correlation between the final temperature ratio $x_f$ and $\chi^2$ for noncanonical (left) and canonical (right) models from the MCMC analysis, that 
    respect the  95\% C.L.\ limits of \textit{Planck}. The grey shading at the top indicates the excluded region $x_f>0.5$. For comparison, the vertical lines show the minimal value of $\chi^2$ for single-field $p=1/2$ (green) and $p=2/3$ (orange) inflation models. Data points above the dashed line in the left panel correspond to models where the oscillations of $Y$ are overdamped, as explained in the text.    }
    \label{fig:scatter2}
\end{figure*}

\section{Markov Chain Monte Carlo}
\label{sectIV}

To more completely explore the parameter space, we 
performed a Markov Chain Monte Carlo (MCMC) analysis for both the canonical and noncanonical models. The parameters that were allowed to vary were the initial field amplitudes and the kinetic parameter $f$ in Eq.~(\ref{eq:nmk}), and we ran separate MCMCs for the values $p=1/3,\,1/2,\, 2/3$,  selecting models that fell within the \textit{Planck} 95\% confidence limits.   Scatter plots showing correlations of the input parameter values from the resulting MCMC chains are presented in Fig.~\ref{fig:mcmcparam}.  They demonstrate the requirement of a
small initial asymmetry $Y_i\cong X_i$, and an inverse correlation between $f$
and $X_i$ that is necessary for achieving a long enough period of inflation.  Fig.~\ref{fig:mcmcparam} shows that smaller initial field values are required in the noncanonical models, but in terms of the 
canonically normalized fields, {\it e.g.,} $U = \int F(X)^{1/2}dX$, the initial values are of the same order ($\gtrsim 10$) for both kinds of models.

\subsection{Inflationary observables}

As explained above, canonical and noncanonical models lead to very similar predictions for $n_s$
and $r$, differing primarily in their predicted final temperature ratios.
Fig.~\ref{fig:scatter1} shows these predictions, as well as the correlation between the total duration of inflation $N_{\rm tot}$ and the initial temperature asymmetry $x_i$, for both kinds of models. There is no apparent distinction between the two kinds of models in these plots.  For the value of $N_*$ at which $n_s$ and $r$ were evaluated in each model, we chose a random number in the interval $[50,60]$, to reflect the uncertainty in the overall scale of inflation and reheat temperatures.

We observe a significant variation for the value of the spectral index (and, to a lesser extent, for the tensor-to-scalar ratio) within each class of models depending on the parameters and initial conditions. This is the result of the dependence of $n_s$ on both $N_*$ and the field ratio $\xi_*$ at horizon crossing described by Eq.~(\ref{eq:nsapprox}), and it demonstrates the versatility of the
two-field models in contrast to single-field inflation.

The right panel of Fig.~\ref{fig:scatter1} can be understood in terms of the CTHC
effect described previously:
if the fields are close to each other at the beginning of inflation ($x_i=\xi_i^{p/4}\approx 1$), then inflation must last longer in order for the trajectory to bend enough to reach the ratio $\xi_*=(Y_*/X_*)^{2/p}$ at horizon crossing that corresponds to optimal agreement with CMB observations. One can achieve a longer period of slow-roll either by increasing the initial values of the fields or the nonmiminal kinetic parameter $f$. Numerically, we find that
\be
0.4\lesssim\xi_*\lesssim0.7
\label{optrange}
\ee
is required at horizon exit for $n_s$ to fall within \textit{Planck}'s 68\% confidence limit, $n_s=0.9649\pm0.0042$ \cite{Akrami:2018odb}, although the precise interval depends mildly on $p$. 

It is noteworthy that for a given value of $p$, there exists a minimum value
of $x_i$, below which no models were retained in our MCMC chains. This can be
understood in terms of the optimal range (\ref{optrange}) for $\xi_*\cong x_*^{p/4}$.
If $x_i$ falls below the corresponding lower bound of Eq.~(\ref{optrange}),  $\xi_i\lesssim0.4^{p/4}$, then
$x(t)$ continues to 
decrease during inflation, resulting in a value of $n_s$ that is too small no matter how long inflation lasts, in conflict with \textit{Planck} data. These  minimum values are rather close to 1: $x_i\gtrsim 0.86$, $0.92$ and $0.94$ for $p=2/3$, $1/2$ and $1/3$, respectively. Therefore, the flatter the potential is, the closer the energy densities must be initially.  Whether this could be considered as a mild fine tuning
of initial conditions cannot be quantified in the absence of a complete 
theory for the probability distribution of the field amplitudes at the beginning of inflation.  It is surprising and somewhat ironic that our search for a large temperature asymmetry reveals the requirement of a moderate
level of symmetry in the initial conditions, to achieve agreement with current cosmological observations.

\subsection{Temperature ratio}

It is striking that  our mechanism can produce final temperature asymmetries as great as $x_f \sim 10^{-10}$ (for the noncanonical models), and it rarely gives values $x_f \gtrsim 0.5$ that are
in the cosmologically excluded region.  This is illustrated in 
Fig.~\ref{fig:scatter2}, which shows the correlation between  $x_f$  and the value of $\chi^2$ with respect to $n_s$ and $r$. The most important feature is the difference of scale between the vertical axes of the two panels: the noncanonical models can lead to dramatically lower $x_f$ than the canonical ones.

The pertinent difference between the two kinds of models arises from the equation of state of the fields near the minimum of the potential. Since we assume the canonical potential retains its fractional power-law shape in the oscillatory regime, the energy density of the mirror inflaton $Y$ redshifts as in Eq.~(\ref{eq:rhoscaling}) once it starts oscillating. One can see that higher values of $p$ make $Y$ decay faster, which is why $p=2/3$ canonical models may lead to a temperature ratio as low as $x_f\sim10^{-5}$ while $p=1/3$ models are limited to $x_f\sim 10^{-3}$ at best.   These conclusions could change if the shape of the potential was different near its minimum.\footnote{For instance, axion monodromy models \cite{McAllister:2008hb} typically have a potential of the form $V(X)\sim (X^{2}+\epsilon^{2})^{p/2}$, which looks like a power-law when $X\gg\epsilon$ while being approximately quadratic near the minimum, when $X\ll\epsilon$. This scenario would be very similar to our noncanonical model since they would both yield the same equation of state at small field amplitude. }
In contrast, the noncanonical models have a  quadratic potential at small field values, leading to same equation of state as cold dark matter after the end of inflation: $\rho\sim a^{-3}$. Hence in these models the mirror inflaton generally decays much faster than in canonical models, explaining why the final temperature ratio can be as low as $\sim 10^{-10}$.

Exceptionally, the left panel of Fig.~\ref{fig:scatter2} reveals some noncanonical models that lead to small values of $\chi^2$, yet with much larger temperature ratios. These points are separated from the rest by a dashed line to highlight the distinct correlations. They correspond to models where the nonminimal kinetic parameter $f$ is small ($\lesssim 1$, \textit{cf}. the lower part in the second panel of Fig.~\ref{fig:mcmcparam}). Such values of $f$ generally require larger initial amplitudes $X_i$ and $Y_i$ to maintain the requirement (\ref{optrange}) on $\xi_*$. This in turn makes the Hubble parameter larger, increasing the impact of damping on the oscillations of $Y$ in its oscillating phase.

Elaborating on this point, by neglecting the nonminimal kinetic term after $Y$ leaves the slow rolling regime, the equation of motion of the mirror inflaton is
\be 
Y'' + 3 Y' + \frac{m^2}{H^2} Y \approx 0,
\ee 
which describes a damped harmonic oscillator. When $H/m>2/3$, the oscillations of $Y$ are overdamped, making its amplitude decay very slowly. By contrast, if $H/m<2/3$ when $Y$ enters its oscillating phase it is underdamped and it decays exponentially. Models above the dashed line in the right panel of Fig.~\ref{fig:scatter2} all correspond to the overdamped case, which is why the final temperature ratio is many orders of magnitude above the low-amplitude cases. This behavior occurs only in the noncanonical models because there both $f$ and the initial conditions $X_i,Y_i$ impact the duration of inflation; in canonical scenarios only the latter play a role.

\subsection{Predictions for $x_f$ and isocurvature}
Although the final temperature ratios $x_f$ span many orders of 
magnitude in the MCMC results, if future CMB measurements converge
on values of $n_s$ and $r$ near the {\it Planck} best-fit values,
this range can become narrower. 
This is due to the correlation between the spectral index and the field ratio given by Eq.~(\ref{eq:nsapprox}), which implies a correlation between $x_f$ and $n_s$. Hence only certain values of $x_f$ are predicted by models that yield $n_s=0.9649\pm0.0042$. 
The right panel of 
Fig.~\ref{fig:scatter2}, which refers to the canonical models, 
predicts {\it e.g.} $\sim 10^{-3}$--$10^{-2}$ for $p=1/3$.  Wider intervals are allowed 
for the underdamped noncanonical models (below the dashed line of the left panel), {\it e.g.,} 
 the data points cluster near $\chi^2\approx 0.6$ for $x_f\sim10^{-10}$--$10^{-6}$ for $p=1/3$, while the possibility of
 overdamped models extends the most likely range of $x_f$ to up to $\sim 10^{-2}$.

Fig.~\ref{fig:scatter2} shows, by the vertical lines, the minimal values of $\chi^2$ obtainable in single-field $p=2/3$ and $p=1/2$ inflationary model.  Neither model gives a good fit to the data, and the $p=1/3$ case is disfavored at $>95\%$\,C.L. (see Fig.~\ref{fig:nsr}), so its corresponding horizontal line falls outside the plotted region. The single-field models also describe two-field inflation in the regime where $Y\ll X$ at horizon crossing, so that $x_f$ can be arbitrarily small in those scenarios. However, most  points from our MCMC analysis give better fits than the single-field models.

\begin{figure}[t!]
    \centering
    \includegraphics[width=\linewidth]{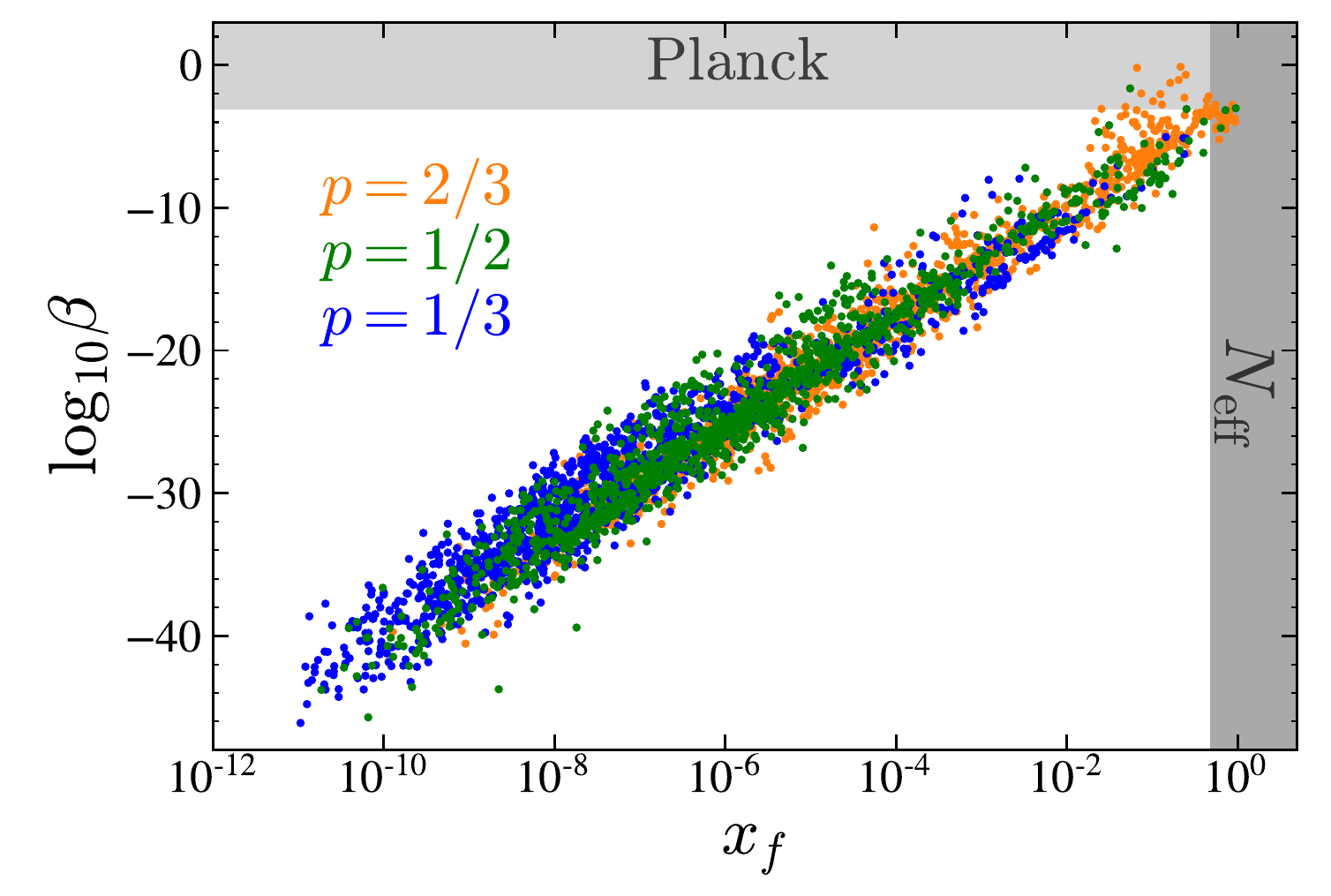} 
    \caption{Correlation between the amplitude of the isocurvature spectrum $\beta$ on large scales and the final temperature ratio $x_f$ in the noncanonical models. The grey bands are the excluded region $x_f>0.5$ and the limit on the \textit{Planck} sensitivity $\beta>10^{-3}$.}
    \label{fig:isocurv}
\end{figure}

Similarly to the benchmark models of section \ref{sec:benchmark}, we computed the amplitude of the isocurvature power spectrum $\beta$ for noncanonical models in the MCMC chains. Fig.\ \ref{fig:isocurv} shows the resulting correlation between $\beta$ and $x_f$. As anticipated, the $\beta$ is too small to be observable in most cases, with only a handful of $p=2/3$ models approaching the sensitivity of \textit{Planck}. We empirically observe a scaling relation $\beta\sim x_f^4$.  Even though models with observable isocurvature are rare,
these examples are interesting because they suggest that a detection of isocurvature could be correlated with an observable
deviation in $N_{\rm eff}$.

\section{Conclusion}
\label{sectV}
In this work we have revisited a mechanism for generating a cooler mirror sector from inflation, due to random initial conditions, while maintaning exact mirror symmetry at the Lagrangian level. It was shown that two-field inflation models with a fractional power-law potential can efficiently amplify a small asymmetry between the visible sector and its mirror counterpart, leading to a temperature ratio as low as $x_f\sim10^{-10}$ at the end of inflation. Models with a nonminimal kinetic term generally lead to values of $x_f$ that are much lower than the corresponding canonically normalized scenarios, due to the difference in the equation of state of the fields when they are oscillating around the minimum of the potential.

While both kinds of models can be in much better agreement with \textit{Planck} data than single-field inflation, somewhat surprisingly this requires  the initial temperature asymmetry between the two sectors to be relatively small, $x_i\gtrsim0.86$.
The value of the tensor-to-scalar ratio $r$ depends on the effective power $p$ of the potential, lying in the range $\sim0.03$--$0.07$ for $p$ between $1/3$ and $2/3$, which could be observed in upcoming CMB experiments \cite{Hazumi:2019lys}.

A key assumption is that the two sectors are decoupled or very weakly interacting with each other.   For example, a $\lambda_{HH'} |H|^2|H'|^2$ coupling mixing the Higgs bosons of the two sectors must have $\lambda_{HH'}\lesssim 10^{-8}$
to avoid equilibration of the temperatures after reheating.\footnote{By demanding the scattering rate $\Gamma \sim \lambda_{HH'}^2 T < H\sim T^2/M_p$ down to the weak scale $T\sim m_H$. }\ \  This is technically natural since $\lambda_{HH'}$ is only multiplicatively renormalized.

On the other hand we have also assumed the possible 
interaction $\lambda_{XY}X^2 Y^2$ to be absent.  One might expect that its presence could synchronize
the two fields during inflation and make $X\approx Y$ at the onset of reheating.   In a preliminary investigation we find
the opposite behavior: nonvanishing $\lambda_{XY}$ instead tends to enhance the final temperature asymmetry, naively estimated as we have done throughout this work.  However
whether this would be a good estimate in the present case
is questionable, because of the possibility of $Y$ particle productions via parametric
resonance or $XX\to YY$ scattering during reheating.  This question is beyond the scope of the present work,
but could be interesting for future study.


\smallskip
{\bf Acknowledgment.}  We thank Eva Silverstein for helpful correspondence.  This work was supported by 
NSERC (Natural Sciences and Engineering Research Council, Canada). \\

\bibliography{ref}{}
\bibliographystyle{elsarticle-num}

\end{document}